\documentstyle[art10]{article}
\begin{document}

\begin{titlepage}

\title{Octonions: Invariant Leech Lattice Exposed }

\author{Geoffrey Dixon\thanks{supported in part by my chair.}
\\  Department of Mathematics or Physics \\
 Brandeis University \\
Waltham, MA 02254 \\
 email: dixon@binah.cc.brandeis.edu \\
\and Department of Mathematics \\
University of Massachusetts \\
Boston, MA 02125 \\
 email: dixon@umbsky.cc.umb.edu}

\maketitle

\begin{abstract}
The structure of a previously developed representation of the Leech
lattice, $\Lambda_{24}$, is exposed to further light with this
unified and very simple construction.
\end{abstract}

\end{titlepage}

\section*{1. Introduction.}

In  {\bf [1]} I presented a representation of $\Lambda_{24}$ (see
{\bf [2]}) over ${\bf O}^{3}$ (octonionic 3-space).   The octonion
multiplication I am accustomed to using (see {\bf [3-7]}) is one of
four for which index cycling and doubling are automorphisms; and the
representation of $\Lambda_{24}$ developed in {\bf [1]} is invariant
under both kinds of maps.

The purpose of the present paper is to give this representation an
elegant characterization that highlights the invariances and the
relationship to the octonion algebra.

As usual, I'm going to ask the interested reader to look at {\bf
[4-7]} for background material on the relationship of the octonions to
$E_{8}$ and $\Lambda_{16}$.  The notation I employ here, and in those
previous papers, is the same I employed in {\bf [3]}.

\section*{2. Foundation.}

Let $X$ be an
arbitrary octonion of unit norm.  Let $A,B \in {\bf O}$.  Then in
general,
%
%1
%
\begin{equation}
A\circ_{X}B \equiv (AX)(X^{\dagger}B) = (A(BX))X^{\dagger} =
X((X^{\dagger}A)B).
\end{equation}
This defines the octonion X-product {\bf [3,4,8]}.

The octonion multiplication I employ is characterized by the following
rule cyclic in the indices:
%
%2
%
\begin{equation}
e_{a}e_{a+1}=e_{a+5}, \; \; a = 1,...,7
\end{equation}
(indices in (2) taken from 1 to 7, modulo 7) (see {\bf [3,4]}.
The index doubling invariance of the resulting multiplication leads
to the following equivalent rules:
$$
e_{a}e_{a+2}=e_{a+3},
$$
and
$$
e_{a}e_{a+4}=e_{a+6}.
$$

Define:
%
%3
%
\begin{equation}
\begin{array}{cl}
\Xi_{0} = & \{\pm e_{a}\},  \\ \\
\Xi_{1} = & \{(\pm e_{a}\pm e_{b})/\sqrt{2}: a,b \mbox{
distinct}\},  \\ \\
\Xi_{2} = & \{(\pm e_{a}\pm e_{b}\pm e_{c}\pm e_{d})/2: a,b,c,d
\mbox{ distinct},   \\ \\
&e_{a}(e_{b}(e_{c}e_{d}))=\pm 1\}, \\ \\
\Xi_{3} = & \{(\sum_{a=0}^{7}\pm e_{a})/\sqrt{8}:
\mbox{ odd number of +'s}  \},   \\ \\
& a,b,c,d\in\{0,...,7\}.  \\
\end{array}
\end{equation} \\
Let $X \in \Xi_{m}$, for some $m=0,1,2,3$.  Given the multiplication
(2), for all $a,b \in \{0,...,7\}$, there is some $c \in \{0,...,7\}$
such that
%
%4
%
\begin{equation}
e_{a}\circ_{X}e_{b}=\pm e_{c}
\end{equation}
(see {\bf [4]}).
Therefore, by (1),
$$
\begin{array}{c}
e_{a}(e_{b}X) = \pm e_{c}X, \\ \\
(X^{\dagger}e_{a})e_{b} = \pm X^{\dagger}e_{c}.
\end{array}
$$
By induction this implies that for all $a,b,...,c \in \{0,...,7\}$,
there exists some $d \in \{0,...,7\}$, such that if $X \in \Xi_{m}$,
for some $m=0,1,2,3$, then
%
%5
%
\begin{equation}
\begin{array}{c}
e_{a}(e_{b}(...(e_{c}X)...)) = \pm e_{d}X, \\ \\
(...((X^{\dagger}e_{a})e_{b})...)e_{c} = \pm X^{\dagger}e_{d}.
\end{array}
\end{equation}

Define
%
%6
%
\begin{equation}
\begin{array}{lll}
{\cal E}_{8}^{even} &=& \Xi_{0}\cup \Xi_{2}, \\ \\
{\cal E}_{8}^{odd} &=& \Xi_{1}\cup \Xi_{3}. \\ \\
\end{array}
\end{equation} \\
These are the inner shells (normalized to unity) of $E_{8}$ lattices
{\bf [5]}.

\section*{3. $\Lambda_{24}$.}

Let $P \in {\cal E}_{8}^{even}$, and define
%
%7
%
\begin{equation}
\ell_{0} = \frac{1}{2}(1+e_{1}+e_{2}+e_{3}+e_{4}+e_{5}+e_{6}+e_{7}).
\end{equation} \\
Define the subset of ${\bf O}^{3}$,
%
%8
%
\begin{equation}
\begin{array}{ccl}
\Lambda_{24}^{3} &=& \{<\frac{1}{2}P, \; \; \pm \frac{1}{2}e_{a}P,
\; \; \pm \frac{1}{2}(e_{a}\ell_{0}e_{a})(e_{b}P)>: \; \; a,b\in
\{0,...,7\}\\ \\
&\cup&\{\mbox{all permutations of such elements}\}, \\
\end{array}
\end{equation} \\
where the two $\pm$'s are independent.

Let
$$
Q=\pm e_{a}P\in {\cal E}_{8}^{even} \Longrightarrow
P = \mp e_{a}Q.
$$
Also, by (5) there exists some index $d\in\{0,...,7\}$ such that
$$
e_{b}P = \mp e_{b}(e_{a}Q) = \pm' e_{d}Q.
$$
Therefore permuting the first two terms of the octonion triple in
(8) leaves the form of that element intact.  As a consequence there
are only three inequivalent permutations on that element, each
characterized by the position of the component with the term
$\ell_{0}$.

Given that there are three inequivalent permutations on the triple
in (8), 16 values for $\pm e_{a}$, 16 for $\pm e_{b}$, and 240
possible values $P$, the order of $\Lambda_{24}^{3}$ is therefore
$$
3\times 16\times 16 \times 240 = 184320.
$$ \\ \\

The term $(e_{a}\ell_{0}e_{a})(e_{b}P)$ can be written in other
ways.  Using (5) and the Moufang identities it is not hard to prove
that there are indices $c,d\in\{0,...,7\}$ such that
%
%9
%
\begin{equation}
\begin{array}{ccl}
(e_{a}\ell_{0}e_{a})(e_{b}P) &=& \pm e_{a}(\ell_{0}(e_{c}P)) \\ \\
&=& \pm' (e_{a}\circ_{P}(\ell_{0}\circ_{P}e_{d}))P.
\end{array}
\end{equation} \\
Therefore the triple in (8) may be written
%
%10
%
\begin{equation}
\frac{1}{2}<1, \; \; \pm e_{a},
\; \; \pm e_{a}\circ_{P}(\ell_{0}\circ_{P}e_{d})>P.
\end{equation} \\
Note that the element $P$ is not completely factored out, since it
is part of the $\circ_{P}$ product in the third component. \\

Finally define
%
%11
%
\begin{equation}
\begin{array}{lll}
\Lambda_{24}^{1} &=& \{<A,0,0>,\; <0,A,0>,\; <0,0,A>: \; \; A \in
{\cal E}_{8}^{even}\},
\end{array}
\end{equation} \\
of order $3\times 240$;
and
%
%4
%
\begin{equation}
\begin{array}{lll}
\Lambda_{24}^{2} &=& \{<A,B,0>,\; <0,A,B>,\; <B,0,A>: \; \;
 A,B \in \frac{1}{\sqrt{2}}{\cal E}_{8}^{odd}, \\ \\
&& AB^{\dagger} =\pm\frac{1}{2}e_{a}, \; \; a\in \{0,...,7\}\},
\end{array}
\end{equation} \\
of order $3\times 16 \times 240$.
Together with $\Lambda_{24}^{3}$ these three sets form the inner
shell of a representation of $\Lambda_{24}$ normalized to unity
(also see {\bf [1]}).  There are a total of
$$
3 \times 240 \; + \; 3 \times 16 \times 240 \; + \;
3 \times 16^{2} \times 240 = 196560
$$
elements in this inner shell.

\section*{4. Invariance.}

Both $\Lambda_{24}^{1}$ and $\Lambda_{24}^{2}$ are easily seen to be
invariant under the index doubling and cycling maps.  That
$\Lambda_{24}^{3}$ is also invariant rests on the invariance of
$$
\ell_{0} = \frac{1}{2}(1+e_{1}+e_{2}+e_{3}+e_{4}+e_{5}+e_{6}+e_{7}).
$$
In fact, it is easy to see that any element of {\bf O} of the form
$$
u+v(e_{1}+e_{2}+e_{3}+e_{4}+e_{5}+e_{6}+e_{7}),
$$
$u,v$ real, will be index doubling and cycling invariant given that
the underlying {\bf O} multiplication is index doubling and cycling
invariant. \\ \\ \\ \\
I'd like to acknowledge several electronic conversations with Tony
Smith, who maintains a fascinating Web site at Georgia Tech:
www.gatech.edu/tsmith/home.html

\end{document}